\documentclass[12pt]{article}

\ifx\pdfoutput\undefined
\usepackage[dvips,bookmarks]{hyperref}
\else
\usepackage{hyperref}
\fi
\hypersetup{colorlinks=false,bookmarksopen,bookmarksnumbered,citecolor=blue,
   pdfstartview=FitH}

\usepackage{latexsym}
\usepackage{amssymb,amsfonts,amsmath}
\usepackage{graphicx} 
\usepackage{indentfirst}
\usepackage{bbm}
\usepackage{amssymb}
\usepackage{verbatim}
\usepackage{amsmath, amsthm,amssymb}
\usepackage{mathrsfs}
\usepackage{hyperref}
\usepackage{amsfonts}
\usepackage{dsfont}

\parskip=\medskipamount

\newcommand {\cA}{{\cal A}}
\newcommand {\cB}{{\cal B}}
\newcommand {\cC}{{\cal C}}
\newcommand {\cD}{{\cal D}}
\newcommand {\cE}{{\cal E}}
\newcommand {\cF}{{\cal F}}
\newcommand {\cG}{{\cal G}}
\newcommand {\cH}{{\cal H}}

\newcommand {\cL}{{\cal L}}

\newcommand {\cN}{{\cal N}}
\newcommand {\cO}{{\cal O}}

\newcommand {\cV}{{\cal V}}
\newcommand {\cW}{{\cal W}}

\def\a{\alpha}
\def\b{\beta}
\def\c{\chi}
\def\d{\delta}
\def\e{\epsilon}

\def\g{\gamma}
\def\G{\Gamma}

\def\m{\mu}

\def\q{\theta}

\def\t{\tau}

\def\D{\Delta}

\def\L{\Lambda}

\def\S{\Sigma}

\def\tr{{\rm tr}}

\def\ri{{\rm i}}
\def\re{{\rm e}}

\newcommand{\ad}{{\dot{\alpha}}} 
 
\newcommand{\ve}{\varepsilon}

\newcommand{\hf}{\frac12}

\newcommand{\be}{\begin{equation}}
\newcommand{\ee}{\end{equation}}
\newcommand{\bea}{\begin{eqnarray}}
\newcommand{\eea}{\end{eqnarray}}
\newcommand{\non}{\nonumber}
\newcommand{\ba}{\begin{array}}
\newcommand{\ea}{\end{array}}

\newcommand{\dsR}{{\mathbb R}}

\def\double #1{#1{\hbox{\kern-2pt $#1$}}}

\newcommand{\ha}{{\hat{a}}}
\newcommand{\hb}{{\hat{b}}}
\newcommand{\hc}{{\hat{c}}}
\newcommand{\hd}{{\hat{d}}}
\newcommand{\he}{{\hat{e}}}

\newcommand{\hal}{{\hat{\a}}}
\newcommand{\hbe}{{\hat{\b}}}
\newcommand{\hga}{{\hat{\g}}}

\newcommand{\bsubeq}{\begin{subequations}}
\newcommand{\esubeq}{\end{subequations}}

\newcommand{\eps}{{\ve}}

\newcommand{\rd}{\mathrm d}
\numberwithin{equation}{section}

\renewcommand{\eps}{\ve}

\begin{document}
\begin{titlepage}
\begin{flushright}
September, 2013\\
\end{flushright}

\begin{center}
{\Large \bf 
On supersymmetric  Chern-Simons-type theories \\
in five dimensions
}
\end{center}

\begin{center}
{\bf
Sergei M. Kuzenko
and
Joseph Novak
} \\
\vspace{5mm}

\footnotesize{
{\it School of Physics M013, The University of Western Australia\\
35 Stirling Highway, Crawley W.A. 6009, Australia}}  
~\\
\texttt{joseph.novak@uwa.edu.au}\\
\vspace{2mm}

\end{center}

\begin{abstract}
\baselineskip=14pt
We present a closed-form expression for the supersymmetric non-Abelian Chern-Simons action
in conventional five-dimensional $\cN=1$ superspace. Our construction 
makes use of the superform formalism to generate supersymmetric invariants. 
Similar ideas are applied to construct supersymmetric actions for off-shell supermultiplets 
with an intrinsic central charge.  In particular, the large tensor supermultiplet is described in superspace
for the first time. 
\end{abstract}

\vfill
\end{titlepage}

\newpage
\renewcommand{\thefootnote}{\arabic{footnote}}
\setcounter{footnote}{0}

\tableofcontents{}
\vspace{1cm}
\bigskip\hrule


\section{Introduction}

In the 1990s, it was demonstrated \cite{Witten,Seiberg,IMS} that in five dimensions (5D)
a Chern-Simons term is generated in a supersymmetric Yang-Mills theory by integrating 
out massive hypermultiplets and keeping only the gauge field of the vector supermultiplet. 
In a manifestly supersymmetric setting, which takes into account the entire vector 
supermultiplet, a related one-loop calculation 
was given in \cite{K2006}, both in the Coulomb and non-Abelian phases. 
Using the covariant harmonic supergraphs  \cite{BBKO} and the heat kernel 
techniques in harmonic superspace \cite{KM}, it was shown \cite{K2006} that the hypermultiplet
effective action contains a supersymmetric Chern-Simons (SCS)  term.  

Within the component approach, the off-shell non-Abelian SCS action 
(in the presence of 
conformal supergravity)
in five dimensions  was first constructed by Kugo and Ohashi \cite{KO}. 
Their approach, however, was not systematic. 
They started with the Abelian SCS action,\footnote{The off-shell Abelian SCS action
in five dimensions was constructed for the first time by Zupnik in 5D $\cN =1$
harmonic superspace \cite{Zupnik99}.}
which was efficiently derived using the linear 
supermultiplet action, and then extended it by adding appropriate non-Abelian structures, 
order by order in the coupling constant, 
in such a way as to make the action supersymmetric. 
In the flat space limit, the non-Abelian SCS action is superconformal, 
which makes this theory very interesting for various applications.

Unlike the component construction of \cite{KO}, 
a closed-form expression for the non-Abelian SCS action has never been given
in a superspace setting.  In the Abelian case, the SCS action was derived in 
the 5D $\cN=1$  harmonic \cite{Zupnik99} and 
projective  \cite{KL} superspaces,\footnote{The relationship 
between the 4D $\cN=2$  harmonic \cite{GIKOS,GIOS}
and projective \cite{KLR,LR-projective,LR-projective2} superspace formulations
is spelled out in \cite{K-double} (see also \cite{Kuzenko:2010bd} for a recent review).
The same relationship holds in the case of 5D $\cN=1$ supersymmetry.} 
and also in terms of 4D $\cN=1$ superfields \cite{A-HGW}.\footnote{The action 
given in \cite{A-HGW} was derived using an ad hoc procedure;
this action is trivially deduced from the systematic projective-superspace construction of \cite{KL}.} 
As concerns the non-Abelian case, there exists a unique definition  \cite{Zupnik99}
for the variation of the SCS action with respect an infinitesimal deformation of the analytic gauge prepotential, 
$V^{++} \to V^{++} +\d V^{++}$, 
which describes the Yang-Mills supermultiplet 
within the harmonic-superspace approach.\footnote{The one-loop 
calculation in \cite{K2006} consisted of demonstrating 
that varying the hypermultiplet effective action produces a  SCS action \cite{Zupnik99}
as the leading quantum correction.} However, it is not yet known how to integrate this 
variation in a closed form (see the erratum to \cite{Zupnik99}).  In the projective-superspace
approach,  the variation of the non-Abelian SCS action can be defined similar 
to \cite{Zupnik99} using the formalism of \cite{KL}. But it is also unclear how to integrate it. 
For completeness, it is worth mentioning  
the attempt to construct a non-Abelian SCS action in terms 
of 4D $\cN=1$ superfields \cite{Ziegler,HM-RZ}.
But their action is valid only in a Wess-Zumino gauge, and therefore
it is hardly useful. 

In this paper we present a 
closed-form 
expression for the non-Abelian SCS action
in the conventional 5D $\cN=1$ superspace setting described in \cite{KL}. 
To achieve this, we do not define the action as an integral over the superspace 
or its analytic subspace. Instead, we 
adopt the superform
construction of supersymmetric invariants
\cite{Castellani,Hasler,Ectoplasm,GGKS},
also known as the rheonomy approach \cite{Castellani} or the ectoplasm 
approach \cite{Ectoplasm,GGKS}.
More specifically, we will build on 
the recent papers \cite{KT-M13, BKNT-M2,KNT-M} 
in which $\cN\leq 6$ conformal supergravity 
actions in three dimensions have been constructed efficiently and elegantly 
via the superform 
approach by making use of the Chern-Simons form together with a curvature induced form. 
This method is a generalization of the superform 
formulation for the linear supermultiplet in four-dimensional $\cN = 2$ conformal supergravity given in \cite{BKN}.\footnote{This is  an example of a known construction where an invariant derived from a closed super
$d$-form can be generated from a closed, gauge-invariant super $(d+1)$-form
provided that the latter is Weil trivial, i.e. exact in invariant cohomology
(a concept introduced by Bonora, Pasti and Tonin \cite{BPT} in the
context of anomalies in supersymmetric theories). Examples of this include
Green-Schwarz actions for various branes \cite{HRS}, as well as some
higher-order invariants in other supersymmetric theories which were studied, e.g.,
 in \cite{BHLSW,BHS13}.} 
Such an approach can be adapted to five-dimensions 
 and we endeavor to demonstrate this for the non-Abelian SCS theory.

The superform approach can also be used to describe the dynamics
of 5D off-shell supermultiplets with an intrinsic central charge. Of course, 
such theories have been studied in components \cite{KO,KO-I,KO-III} 
and in harmonic superspace
\cite{KL,Kuzenko:5D}. In the component setting, however, one has to use rather different ideas 
in order to describe (i)  the non-abelian Chern-Simons theory and (ii) 
the models for off-shell supermultiplets with an intrinsic central charge.
As will be shown below, if the superform approach is employed the two types of theories are  formulated uniformly 
in superspace.

This paper is organized as follows. In section \ref{NACST}, we introduce the superform formulation of the Yang-Mills supermultiplet and use it to 
construct the Chern-Simons action. To do so we use both the Chern-Simons form and a curvature induced form that we will introduce. 
In section \ref{MWCC}, we turn to supermultiplets with central charge. 
We provide both the superform formulations for a gauge two-form supermultiplet and the linear supermultiplet with central charge. This immediately leads to  
the action principle based on the linear supermultiplet. 
Concluding comments are given 
in section \ref{conclusion}. Finally, in the appendix we analyze the possibility to have 
a gauge connection that is not annihilated by the central charge. 

Throughout the paper, we follow the 5D notation and conventions of \cite{KL}.


\section{Non-Abelian Chern-Simons theory} \label{NACST}

In this section, we describe the non-Abelian SCS theory based on a Yang-Mills supermultiplet 
and derive the corresponding action via the superform approach.


\subsection{Yang-Mills supermultiplet} \label{YM}

Conventional 5D $\cN=1$ Minkowski superspace $\dsR^{5|8}$ may be parametrized 
by the coordinates $z^{\hat{A}} = (x^{\hat{a}}, \theta^{\hat{\a}}_i)$. One can introduce flat covariant derivatives 
$D_{\hat{A}} = (\partial_{\hat{a}}, D_{\hat{\a}}^i)$ which obey the 
algebra
\begin{align}
[ D_{\hat{A}}, D_{\hat{B}} \} &= T_{\hat{A} \hat{B}}{}^{\hat{C}} D_{\hat{C}} \ ,
\end{align}
with
\be T_{\hat{\a}}^i{}_{\hat{\b}}^j{}^{\hat{c}} = - 2 \ri \eps^{ij} (\G^{\hat{c}})_{\hat{\a} \hat{\b}}
\ee
the only non-vanishing torsion component.

The non-Abelian vector supermultiplet may be described in superspace by introducing the gauge covariant derivatives\footnote{Keep in mind that the operation of complex conjugation 
acts as $(D_{\hat{\a}}^i F)^* = - (-1)^{\eps(F)} D^{\hat{\a}}_i F^{*}$, where $\eps(F)$ is the Grassmann parity of $F$, see \cite{KL} for details.} 
\be \cD_{\hat{A}} = (\cD_{\hat{a}} , \cD_{\hat{\a}}^i) = D_{\hat{A}} + \ri V_{\hat{A}}(z) \ , \quad \{ \cD_{\hat{A}} , 
\cD_{\hat{B}} \} = T_{\hat{A} \hat{B}}{}^{\hat{C}} \cD_{\hat{C}} + \ri F_{\hat{A} \hat{B}} \ , \label{CovDerAlgebra}
\ee
where $V_{\hat{A}}$ is a gauge connection taking values in the Lie algebra of the gauge group. The covariant derivatives and field strength may also be written in a 
coordinate-free way as follows
\be \cD = \rd + \ri V \ , \quad F = \rd V - \ri V \wedge V \ , \label{CovDer}
\ee
where
\be \cD := \rd z^{\hat{A}} \cD_{\hat{A}} \ , \quad V := \rd z^{\hat{A}} V_{\hat{A}} \ , \quad F := \hf \rd z^{\hat{B}} \wedge \rd z^{\hat{A}} F_{\hat{A} \hat{B}} \ .
\ee

The covariant derivatives 
possess the gauge transformation law
\be \cD_{\hat{A}} \rightarrow \re^{\ri \t} \,\cD_{\hat{A}} \, \re^{- \ri \t} \ , 
\quad \t^\dag = \t \ , 
\ee
where the Lie-algebra-valued gauge parameter $\t(z)$ is arbitrary 
modulo the reality condition imposed. 
This implies that the gauge connection and field strength transform as follows
\be
V \rightarrow  \re^{\ri \t } \,V \,e^{- \ri \t} - \ri \re^{\ri \t } \,\rd \re^{- \ri \t } \ , 
\quad F \rightarrow \re^{\ri \t } \,F \,\re^{- \ri \t} \ .
\ee

The field strength satisfies the Bianchi identity
\be \cD F = \rd F + \ri V \wedge F - \ri F \wedge V = 0 \ , \quad \cD_{[\hat{A}} F_{\hat{B} \hat{C} \} } - T_{[\hat{A} \hat{B}}{}^{\hat{D}} F_{|\hat{D}|\hat{C} \} } = 0 \ .
\ee
Upon constraining the lowest mass dimension component of the field strength tensor as 
\cite{HL,Zupnik99,KL}
\bsubeq \label{FSComponents}
\be F_{\hat{\a}}^i{}_{\hat{\b}}^j = - 2 \ri \eps^{ij} \eps_{\hat{\a} \hat{\b}} W \ , \label{YMConstraint}
\ee
the remaining components are found to be
\be  \quad F_{\hat{a}}{}_{\hat{\b}}^j = (\G_{\hat{a}})_{\hat{\b}}{}^{\hat{\g}} \cD_{\hat{\g}}^j W \ , 
\quad F_{\hat{a}\hat{b}} = \frac{\ri}{4} (\S_{\hat{a}\hat{b}})^{\hat{\a} \hat{\b}} \cD^k_{(\hat{\a}} \cD_{\hat{\b}) k} W \ .
\ee
\esubeq
Here the superfield $W$ is Hermitian, $W^\dag = W$, and obeys the superfield Bianchi identity
\be \cD_{\hat{\a}}^{(i} \cD_{\hat{\b}}^{j)} W = \frac{1}{4} \eps_{\hat{\a} \hat{\b}} \cD^{\hat{\g} (i} \cD_{\hat{\g}}^{j)} W \ .
\ee
From the above constraint one can derive identities involving products of spinor derivatives acting on $W$. We list these below:
\bsubeq \label{VMIdentities}
\begin{align}
\cD_{\hat{\a}}^{i} \cD_{\hat{\b}}^j W &= - \hf \eps^{ij} \cD_{(\hat{\a}}^k \cD_{\hat{\b}) k} W + \frac{1}{4} \eps_{\hat{\a} \hat{\b}} \cD^{\hat{\g} (i} \cD_{\hat{\g}}^{j)} W 
- \ri \eps^{ij} \cD_{\hat{\a}\hat{\b}} W 
\ , \\
\cD_{\hat{\a}}^i \cD^{\hat{\g} (j} \cD_{\hat{\g}}^{k)} W &= \frac{1}{3} \eps^{ij} \cD_{\hat{\a} l} \cD^{\hat{\g} (k} \cD_{\hat{\g}}^{l)} W
+ \frac{1}{3} \eps^{ik} \cD_{\hat{\a} l} \cD^{\hat{\g} (j} \cD_{\hat{\g}}^{l)} W \non\\
&= 8 \ri \eps^{i(j} \cD_{\hat{\a} \hat{\g}} \cD^{\hat{\g} k)} W 
- 8 \eps^{i (j} [W,  \cD_{\hat{\a}}^{k)} W] \ , \\
\cD_{\hat{\a}}^i \cD_{(\hat{\b}}^k \cD_{\hat{\g} ) k} W &= 4 \ri \eps_{\hat{\a} (\hat{\b}} \cD_{\hat{\g} ) \hat{\d}} \cD^{\hat{\d} i} W - 4 \ri \cD_{\hat{\a} ( \hat{\b}} \cD_{\hat{\g})}^i W
\ .
\end{align}
\esubeq

As a result of the above identities, we may define the independent fields contained in $W$ as
\begin{align} \label{VMcompDef}
\varphi := W| \ , \quad \Psi_{\hat{\a}}^i := - \ri \cD_{\hat{\a}}^i W | \ , \quad F_{\hat{\a} \hat{\b}} := \frac{\ri}{4} \cD^k_{(\hat{\a}} \cD_{\hat{\b)} k} W | \ , 
\quad X^{ij} := \frac{\ri}{4} \cD^{\hat{\a} (i} \cD^{j)}_{\hat{\a}} W | \ ,
\end{align}
where the bar projection of a superfield $U(z) =U(x,\q)$ is defined by 
the standard rule $U| := U(x,\q)|_{\theta = 0}$. The component gauge field is identified with $V_{\hat{a}}|$ and we will drop the bar projection 
when it is clear that we are referring to the component field. The component field strength $F_{\ha\hb}$ can 
be expressed in terms of the gauge field as follows
\be
F_{\ha \hb} = 2   \partial_{[\ha} V_{\hb]} + \ri \big[V_{\ha} ,V_{\hb } \big] \ .
\ee
It is seen that the vector supermultiplet consists of the following component fields: 
$\varphi$, $\Psi_\hal^i$, $V_\ha$ and $X^{ij}$. 

The supersymmetry transformations of the fields 
$\varphi$, $\Psi_\hal^i$, $V_\ha$ and $X^{ij}$
may be obtained by evaluating 
the component projection of the identities \eqref{VMIdentities}. This gives
\bsubeq \label{VMSUSY}
\begin{align}
\d_\xi \varphi &= \ri \xi^\hga_k \Psi_\hga^k \ , \\
\d_\xi \Psi_\hal^i &= - 2 \xi^{\hbe i} F_{\hal \hbe} + \xi_{\hal j} X^{ij} + \xi^{\hbe i} \cD_{\hbe \hal} \varphi \ , \\
\d_{\xi} X^{ij} &= - 2 \ri \xi^{\hal (i} \cD_\hal{}^\hbe \Psi_\hbe^{j)} - 2 \xi^{\hal (i} [\varphi , \Psi_\hal^{j)}] \ , \\
\d_{\xi} V_\ha &= \xi^\hbe_j F_\hbe^j{}_\ha| = - \ri \xi^\hal_j (\G_\ha)_\hal{}^\hbe \Psi_\hbe^j \ ,
\end{align}
\esubeq
where we have used $\cD_\ha$ to mean its projection, $\cD_\ha| = \partial_\ha + \ri V_\ha|$, when acting on a component field.


\subsection{Superforms and the Chern-Simons action}

The SCS action may readily be found in the Abelian case with the use of the action principle based on a linear supermultiplet without central charge. 
However, a  generalization of the action principle to the non-Abelian case is not straightforward. 
In components, the non-Abelian SCS action was constructed  by Kugo and Ohashi \cite{KO} by first starting with the Abelian 
Chern-Simons action. They added non-Abelian structures to the action and checked supersymmetry with the supersymmetry transformations 
of the non-Abelian theory.

There is a more elegant alternative offered by the superform approach to construct
supersymmetric invariants. 
In conventional 5D superspace $\dsR^{5|8}$, the formalism 
makes use of a closed five-form
\be J = \frac{1}{5!} \rd z^{\hat{E}} \wedge \rd z^{\hat{D}} \wedge \rd z^{\hat{C}} \wedge \rd z^{\hat{B}} \wedge \rd z^{\hat{A}} J_{\hat{A}\hat{B}\hat{C}\hat{D}\hat{E}} \ , \quad 
\rd J = 0 \ .
\ee
Under an infinitesimal coordinate transformation generated by a vector field $\xi = \xi^A \partial_A$, the five-form varies as
\be \d_{\xi} J = \cL_{\xi} J \equiv i_\xi \rd J + \rd i_\xi J = \rd i_\xi J \ .
\ee
If we assume that the components $\xi^A$ vanish at infinity in $\dsR^{5|8}$ then we have the supersymmetric invariant
\bea 
S = \int_{ \dsR^{5}} i^* J~,
\label{ectoS}
\eea
where $i : \dsR^{5} \to \dsR^{5|8} $ is the inclusion map.
This can be represented as 
\bea
S= \int \rd^5 x  \,{}^* J |_{\q=0}\ , \qquad 
 {}^*J = \frac{1}{5!} \eps^{\hat{a}\hat{b}\hat{c}\hat{d}\hat{e}} J_{\hat{a}\hat{b}\hat{c}\hat{d}\hat{e}}  \ .
\label{ectoS2}
\eea

A suitable  action  must also be  invariant under all  {\it gauge symmetries} 
of a dynamical system under consideration. If the closed five-form $J$ 
also transforms by an exact form under the gauge transformations, 
\bea
 \d J =  \rd \Theta \ ,
\eea
then the functional \eqref{ectoS} is a suitable candidate for an action.

For the Chern-Simons action, following \cite{KT-M13, BKNT-M2}, we will construct a gauge invariant closed five-form by first finding 
two solutions, $  \S_{\rm CS}$ and $\S_R$,  to the superform equation
\be
\rd \S = \langle F^3 \rangle := \tr \Big( F \wedge F \wedge F \Big) \ .
\ee
The first of which is the Chern-Simons form $\S_{\rm CS}$. The existence of the second solution, the curvature induced form $\S_R$, is a direct consequence of the 
constraints we imposed on the geometry, eq. \eqref{YMConstraint}. If they transform by an exact form under the gauge group then 
their difference 
\be J = \S_{\rm CS} - \S_{R}
\ee
will yield an appropriate closed five-form that describes the action.


\subsubsection{Chern-Simons five-form}


Representing 
$\langle F^3 \rangle = \rd \, \S_{\rm CS} $ yields the Chern-Simons form
\be \S_{\rm CS} = \tr \Big( V \wedge F \wedge F + \frac{\ri}{2} V \wedge V \wedge V \wedge F - \frac{1}{10} V \wedge V \wedge V \wedge V \wedge V \Big) \ .
\ee
Since $\S_{\rm CS}$ has been constructed by extracting a total derivative from the gauge invariant superform $\langle F^3 \rangle$ it must transform by 
a closed form under the gauge group. In fact, one can show it transforms by an exact form,
\be \S_{\rm CS} \rightarrow \S_{\rm CS} - \rd \ \tr \Big( \rd \t \wedge \big( V \wedge F + \frac{\ri}{2} V \wedge V \wedge V \big) \Big) \ .
\ee


\subsubsection{Curvature-induced five-form}


To construct the curvature-induced five-form we need to find a gauge-invariant solution to
\begin{subequations} \label{CIFSE}
\bea 
\rd \S &=& \tr \Big(F \wedge F \wedge F \Big) \quad 
\eea
or, equivalently,
\bea
2 \cD_{[A} \S_{BCDEF\}} &-& 5 T_{[AB}{}^G \S_{|G|CDEF\}} = 30\, \tr \, 
\Big( F_{[AB} F_{CD} F_{EF \} }\Big) \ , 
\eea
\end{subequations}
where the gauge covariant derivative $\cD_A$ is defined by eq. \eqref{CovDerAlgebra}. 
Note that since $\S$ is a gauge singlet we have
\bea 
\cD_A \S = D_A \S \ .
\eea
Keeping this in mind, we will use gauge covariant derivatives everywhere in this section.

On dimensional grounds, it is natural to impose the constraint\footnote{We denote pairs of spinor and isospinor indices, { \rm e.g.} ${}_{\hat{\a}}^i$ by underlined spinor indices, { \rm e.g.} $\underline{\hat{\a}}$.}
\be \S_{\underline{\hat{\a}} \underline{\hat{\b}} \underline{\hat{\g}} \underline{\hat{\d}} \underline{\hat{\e}}} = 0 \ .
\ee
Then analyzing the superform equation \eqref{CIFSE} by increasing mass dimension and using the identities \eqref{VMIdentities} yields all the remaining 
components of the curvature induced five-form. One 
finds the following components:
\bsubeq
\begin{align} \S_{\hat{a}}{}_{\hat{\a}}^i{}_{\hat{\b}}^j{}_{\hat{\g}}^k{}_{\hat{\d}}^l =& - 4 \Big( \eps^{ij} \eps^{kl} \big( (\G_{\hat{a}})_{{\hat{\a}}{\hat{\b}}} \eps_{{\hat{\g}}{\hat{\d}}} 
+ (\G_{\hat{a}})_{{\hat{\g}}{\hat{\d}}} \eps_{{\hat{\a}}{\hat{\b}}} \big)
+ \eps^{ik} \eps^{jl} \big( (\G_{\hat{a}})_{{\hat{\a}}{\hat{\g}}} \eps_{{\hat{\b}}{\hat{\d}}} + (\G_{\hat{a}})_{{\hat{\b}}{\hat{\d}}} \eps_{{\hat{\a}}{\hat{\g}}} \big) \non\\
&+ \eps^{il} \eps^{jk} \big( (\G_{\hat{a}})_{{\hat{\a}}{\hat{\d}}} \eps_{{\hat{\b}}{\hat{\g}}} + (\G_{\hat{a}})_{{\hat{\b}}{\hat{\g}}} \eps_{{\hat{\a}}{\hat{\d}}} \big)\Big) \tr(W^3) \ , \\
\S_{\hat{a} \hat{b}}{}_{\hat{\a}}^i{}_{\hat{\b}}^j{}_{\hat{\g}}^k =& \ - 4 \ri \Big( \eps^{jk} \eps_{\hat{\b}\hat{\g}} (\S_{\hat{a}\hat{b}})_{{\hat{\a}}}{}^{\hat{\d}} \cD_{\hat{\d}}^i 
+ \eps^{ij} \eps_{{\hat{\a}} \hat{\b}} (\S_{\hat{a}\hat{b}})_{\hat{\g}}{}^{\hat{\d}} \cD_{\hat{\d}}^k 
+ \eps^{ki} \eps_{\hat{\g} {\hat{\a}}} (\S_{\hat{a}\hat{b}})_{\hat{\b}}{}^{\hat{\d}} \cD_{\hat{\d}}^j    \Big) \tr (W^3) \ , \\
\S_{\hat{a} \hat{b} \hat{c}}{}_{\hat{\a}}^i{}_{\hat{\b}}^j  =& - \frac{3}{4} \eps^{ij} \eps_{\hat{\a} \hat{\b}} \eps_{\hat{a} \hat{b} \hat{c} \hat{d} \hat{e}} 
(\S^{\hat{d} \hat{e}})^{\hat{\g} \hat{\d}} \tr (W^2 \cD_{\hat{\g}}^k \cD_{\hat{\d} k} W + 4 W \cD_{\hat{\g}}^k W \cD_{\hat{\d} k} W) \non\\
&- \frac{3}{2} \eps_{\hat{a} \hat{b} \hat{c} \hat{d} \hat{e}} (\S^{\hat{d} \hat{e}})_{\hat{\a} \hat{\b}} \tr ( W^2 \cD^{\hat{\g} (i} \cD^{j)}_{\hat{\g}} W + 4 W \cD^{\hat{\g} (i} W \cD^{j)}_{\hat{\g}} W) \ , \\
\S_{\hat{a} \hat{b} \hat{c} \hat{d}}{}_{\hat{\a}}^i  =& - \frac{\ri} {8} \eps_{\hat{a} \hat{b} \hat{c} \hat{d} \hat{e}}  (\G^{\hat{e}})_{\hat{\a}}{}^{\hat{\b}}
\tr \Big( 6 W \{ \cD^j_{(\hat{\b}} \cD_{\hat{\g}) j} W , \cD^{\hat{\g} i} W \} + 3 W \{ \cD^{\hat{\g} (i} \cD_{\hat{\g}}^{j)} W , \cD_{\hat{\b} j} W \} \non\\
&+ 16 \cD^{\hat{\g} (i} W \cD_{\hat{\g}}^{j)} W \cD_{\hat{\b} j} W \Big) \non\\
& - \ri \eps_{\hat{a} \hat{b} \hat{c} \hat{d} \hat{e}} (\G^{\hat{e}})^{\hat{\b} \hat{\g}} \tr \big( \cD_{\hat{\b}}^{(i} W \cD_{\hat{\g}}^{j)} W \cD_{\hat{\a} j} W \big) \non\\
& + 3 \eps_{\hat{a} \hat{b} \hat{c} \hat{d} \hat{e}} (\S^{\hat{e} \hat{f}})_{\hat{\a}}{}^{\hat{\b}} \tr \Big( W \cD_{\hat{f}} \{ W , \cD_{\hat{\b}}^i W \} \Big) \non\\
& + \frac{3}{2} \eps_{\hat{a} \hat{b} \hat{c} \hat{d} \hat{e}} \tr \Big( W \cD^{\hat{e}} \{ W , \cD_{\hat{\a}}^i W \} \Big)
\ .
\end{align}
The final component
\begin{align} \S_{\hat{a} \hat{b} \hat{c} \hat{d} \hat{e}} =& - \frac{3}{32}  \eps_{\hat{a} \hat{b} \hat{c} \hat{d} \hat{e}} \tr \Big( 
W \cD^{\hat{\g} (k} \cD_{\hat{\g}}^{l)} W  \cD^{\hat{\d}}_{(k} \cD_{\hat{\d} l)} W 
- 2 W \cD^{( \hat{\g} k} \cD^{\hat{\d})}_k W \cD^l_{(\hat{\g}} \cD_{\hat{\d} ) l } W \non\\
& + 4 \cD^{\hat{\g} (k} \cD_{\hat{\g}}^{l)} W \cD^{\hat{\d}}_k W \cD_{\hat{\d} l} W - 8 \cD^{( \hat{\g} k} \cD^{\hat{\d})}_k W \cD^l_{\hat{\g}} W \cD_{\hat{\d} l } W \non\\
& - 16 W \cD^{\hat{f}} \{ W , \cD_{\hat{f}} W \}  + 16 \ri W [ \cD_{\hat{\g} \hat{\d}} \cD^{\hat{\g} k} W,  \cD^{\hat{\d}}_k W ] \non\\
&- 32 W^2 \cD^{\hat{\g} k} W \cD_{\hat{\g} k} W
\Big) \label{md5Comp}
\end{align}
\esubeq
is the most important from the point of view of constructing the action. 
It is obvious that the superform constructed is gauge invariant. 
The last term in \eqref{md5Comp}, which is quartic in $W$, 
 disappears in the Abelian case. 

Once all components are determined there still remains the final superform component equation
\be 5 \cD_{[\hat{a}} \S_{\hat{b} \hat{c} \hat{d} \hat{e}] \underline{\hat{\a}}} - \cD_{\underline{\hat{\a}}} \S_{\hat{a} \hat{b} \hat{c} \hat{d} \hat{e}} 
- 90 \tr \Big( F_{[\hat{a} \hat{b}} F_{\hat{c} \hat{d}} F_{\hat{e} ] \underline{\a}} \Big) = 0 \ .
\ee
However, this only remains as a check as it will always be identically satisfied (see appendix of \cite{Novak1}).


\subsubsection{The component non-Abelian Chern-Simons action}


Making use of the superforms $\S_{\rm CS}$ and $\S_R$ one can construct a closed five-form
\be
J = \S_{\rm CS} - \S_{R} \ ,
\ee
from which one can derive a supersymmetric action. The gauge invariance of the action, modulo total derivatives,  is guaranteed by the fact that 
$\S_{\rm CS}$ transforms via an exact form while $\S_R$ is invariant.

In components we have
\bsubeq
\begin{align} J_{\hat{a}\hat{b}\hat{c}\hat{d}\hat{e}} &=  30 \tr \Big( V_{[\hat{a}} F_{\hat{b}\hat{c}} F_{\hat{d}\hat{e}]} - \ri V_{[\hat{a}} V_{\hat{b}} V_{\hat{c}} F_{\hat{d}\hat{e}]} 
- \frac{2}{5} V_{[\hat{a}} V_{\hat{b}} V_{\hat{c}} V_{\hat{d}} V_{\hat{e}]} \Big)
- \S_{\hat{a}\hat{b}\hat{c}\hat{d}\hat{e}} \ ,
 \end{align}
 or, equivalently, 
 \begin{align}
{}^*J = \frac{1}{4}\eps^{\hat{a}\hat{b}\hat{c}\hat{d}\hat{e}} 
\tr \Big(  V_{\hat{a}} F_{\hat{b}\hat{c}} F_{\hat{d}\hat{e}} - 
\ri V_{\hat{a}} V_{\hat{b}} V_{\hat{c}} F_{\hat{d}\hat{e}} - \frac{2}{5} V_{[\hat{a}} V_{\hat{b}} V_{\hat{c}} V_{\hat{d}} V_{\hat{e}]} \Big) 
- \frac{1}{5!} \eps^{\hat{a}\hat{b}\hat{c}\hat{d}\hat{e}} \S_{\hat{a}\hat{b}\hat{c}\hat{d}\hat{e}} \ .
\label{2.28a}
\end{align}
\esubeq
Applying eq. \eqref{ectoS} to the above results and dividing out an irrelevant factor of $3$ gives the Chern-Simons action
\begin{align}
S 
&= \int \rd^5x \, \tr \Big\{ \frac{1}{12} \eps^{\hat{a}\hat{b}\hat{c}\hat{d}\hat{e}} V_{\hat{a}} F_{\hat{b}\hat{c}} F_{\hat{d}\hat{e}} 
- \frac{\ri}{12} \eps^{\hat{a}\hat{b}\hat{c}\hat{d}\hat{e}} V_{\hat{a}} V_{\hat{b}} V_{\hat{c}} F_{\hat{d}\hat{e}}
-  \frac{1}{30} \eps^{\hat{a}\hat{b}\hat{c}\hat{d}\hat{e}} V_{\hat{a}} V_{\hat{b}} V_{\hat{c}} V_{\hat{d}} V_{\hat{e}} \non\\
&\qquad- \hf \varphi F_{\hat{a} \hat{b}} F^{\hat{a} \hat{b}}+ \hf \varphi X^{ij} X_{ij}
- \frac{\ri}{2} F_{\hat{a} \hat{b}} (\Psi^k \S^{\hat{a} \hat{b}} \Psi_k) \non\\
&\qquad- \frac{\ri}{2} X_{ij} (\Psi^i \Psi^j) + \frac{\ri}{2} \varphi \Psi^k \overleftrightarrow{\! \not \! \! \cD} \Psi_k 
- \varphi \cD^{\hat{a}} \varphi \cD_{\hat{a}} \varphi
- \varphi^2 \Psi^k \Psi_k \Big\} \ ,
\end{align}
where we integrated by parts and defined
\be
\varphi \Psi^k \overleftrightarrow{\! \not \! \! \cD} \Psi_k 
:= \varphi \Psi^k \! \not \! \! \cD \Psi_k - \varphi \! \not \! \! \cD \Psi^k \Psi_k \ .
\ee
The above action may be compared to the action in \cite{KO}. The supersymmetry transformations of the component fields are given by 
eq. \eqref{VMSUSY}.

In the Abelian case the Chern-Simons action simplifies to\footnote{Due to a typo in \cite{KL}, the first term in the 
action differs from the one in \cite{KL} by a factor of $4$.}
\begin{align} \label{abelianVMaction}
S &= \int \rd^5x ~ \tr \Big( \frac{1}{12} \eps^{\hat{a}\hat{b}\hat{c}\hat{d}\hat{e}} V_{\hat{a}} F_{\hat{b}\hat{c}} F_{\hat{d}\hat{e}} 
- \hf \varphi F_{\hat{a} \hat{b}} F^{\hat{a} \hat{b}}+ \hf \varphi X^{ij} X_{ij}
- \frac{\ri}{2} F_{\hat{a} \hat{b}} (\Psi^k \S^{\hat{a} \hat{b}} \Psi_k) \non\\
&\qquad- \frac{\ri}{2} X_{ij} (\Psi^i \Psi^j) + \ri \varphi \Psi^k {\not \! \partial} \Psi_k 
- \varphi \partial^{\hat{a}} \varphi \partial_{\hat{a}} \varphi \Big) \ .
\end{align}
In the next section we will derive the above action with the use of the linear supermultiplet.


\section{Off-shell supermultiplets with central charge} \label{MWCC}

In this section, we provide a superform description for certain supermultiplets with gauged central charge. Firstly, we discuss 
how to gauge the central charge in 5D $\cN=1$ superspace 
following \cite{Kuzenko:5D}.
We then give the superform formulation for the linear supermultiplet 
with central charge and immediately derive the action. Finally, we give the superform formulations for a gauge two-form supermultiplet and for a large tensor supermultiplet.


\subsection{Gauging a central charge in superspace} \label{gauging}

Let $\D$ denote a central charge. It can be gauged using an Abelian vector supermultiplet associated with a gauge connection $\cV$. The procedure 
is similar to the one used in subsection \ref{YM}. We simply need to replace the gauge connection $V$ and field strength 
$F$ in eqns. \eqref{CovDerAlgebra} and \eqref{CovDer} with those associated with the central charge $\D$ as follows:
\be \ri V \rightarrow \cV \D \ , \quad \ri F \rightarrow \cF \D \ .
\ee
The central charge commutes with the covariant derivatives and annihilates both $\cV$ and $\cF$
\be [\D, \cD_{\hat{A}}] = 0 \ , \quad \D \cV = 0 \ , \quad \D \cF = 0 \ .
\ee

Gauge transformations of the covariant derivatives are replaced by
\be \d \cD_{\hat{A}} = [\L \D , \cD_{\hat{A}}] \implies  \d \cV_{\hat{A}} = - D_{\hat{A}} \L \ , \label{CCGTrans}
\ee
where the gauge parameter is inert under the central charge, $\D \L = 0$. The possibility 
of allowing the central charge to not annihilate the gauge connection is discussed in the appendix.

The field strength $\cF$ is constrained to be formally the same as eq. \eqref{FSComponents} but with $W$ replaced by $\cW$. 
For later reference, we list the components of $\cF$ here. They are
\bsubeq \label{VGCC}
\begin{align}
\cF_{\hat{\a}}^i{}_{\hat{\b}}^j &= - 2 \ri \eps^{ij} \eps_{\hat{\a} \hat{\b}} \cW \ , \\
\cF_{\hat{a}}{}_{\hat{\b}}^j &= (\G_{\hat{a}})_{\hat{\b}}{}^{\hat{\g}} \cD_{\hat{\g}}^j \cW \ , \\
\quad \cF_{\hat{a}\hat{b}} &= \frac{\ri}{4} (\S_{\hat{a}\hat{b}})^{\hat{\a} \hat{\b}} \cD^k_{\hat{\a}} \cD_{\hat{\b} k} \cW \ ,
\end{align}
\esubeq
with $\cW$ constrained by the Bianchi identity
\be \cD_{\hat{\a}}^{(i} \cD_{\hat{\b}}^{j)} \cW = \frac{1}{4} \eps_{\hat{\a} \hat{\b}} \cD^{\hat{\g} (i} \cD_{\hat{\g}}^{j)} \cW \ . \label{VGCCBI}
\ee


\subsection{Linear supermultiplet}

Here we construct a superform formulation for 
the 5D linear supermultiplet\footnote{In 4D $\cN=2$ supergravity, 
the linear supermultiplet was introduced by Breitenlohner and Sohnius \cite{BS}
(see also \cite{deWvHVP}) 
building on the rigid supersymmetric construction due to Sohnius 
\cite{Sohnius}. The 5D $\cN=1$ linear supermultiplet  \cite{Zucker} is a natural  
generalization of its 4D ancestor.}
with gauged central charge  
which will naturally lead to the action for the supermultiplet.


\subsubsection{Superform formulation for the linear supermultiplet} \label{SFlinear}

To construct a superform formulation for a supermultiplet with intrinsic central charge one usually makes 
some modifications to superspace. In rigid supersymmetry with a central charge, 
it is well known that one can treat the central charge as a derivative with respect to an additional bosonic 
coordinate. In fact, this approach was used in 4D to construct a superform formulation for the 
linear vector-tensor supermultiplet \cite{HOW, GHH, BHO}.\footnote{The superform formulation and action for the linear supermultiplet with rigid central 
charge in 5D was given in \cite{HPSS-C}. However, the case of a gauged central charge was not studied.} 
For certain supermultiplets, {\it e.g.} the linear supermultiplet, the 
approach is equivalent to dimensional reduction of supermultiplets from higher dimensions. However, the situation is 
more complex in the presence of a gauged central charge. For the linear supermultiplet with gauged central charge in 4D supergravity one 
finds that it is natural to extend the vielbein to include the central charge gauge one-form \cite{BKN}. The resulting formulation turns out to 
be equivalent to a system of superforms. Here we will begin with a generalization of the system of superforms found in \cite{BKN} and 
introduce some useful notation that will help us solve certain constraints.

We introduce a five-form $\tilde{\S}$ and a four-form $\Phi$ which are coupled by the superform equations
\be
\cD \tilde{\S} = \cF \wedge \Phi \ , \quad \cD \Phi = - \D \tilde{\S} \label{SUPERFORMEQNS}
\ee
and transform as scalars under the central charge gauge transformations \eqref{CCGTrans}
\be \d \tilde{\S} = \L \D \tilde{\S} \ , 
\quad \d \Phi = \L \D \Phi \ .
\ee
The superforms $\tilde{\S}$ and $\Phi$ may be related to the linear supermultiplet with central charge by imposing certain 
constraints. It will prove useful to first introduce some notation to deal with the superform equations \eqref{SUPERFORMEQNS}. 

We introduce indices that range over not just $\hat{A}$ but an additional bosonic coordinate, $\hat{\cA} = (\hat{A} , 6)$. Then we may 
rewrite eq. \eqref{SUPERFORMEQNS} in components as
\be \cD_{[\hat{\cA}} \S_{\hat{\cB} \hat{\cC} \hat{\cD} \hat{\cE} \hat{\cF} \}} - \frac{5}{2} T_{[\hat{\cA} \hat{\cB}}{}^{\hat{\cG}} \S_{|\hat{\cG}| \hat{\cC} \hat{\cD} \hat{\cE} \hat{\cF} \} } = 0 \ , \label{BISigma}
\ee
where we have made the identifications
\be T_{\hat{A} \hat{B}}{}^6 = \cF_{\hat{A} \hat{B}} \ , \quad T_{6 \hat{B}}{}^{\hat{\cA}} = T_{\hat{B} 6}{}^{\hat{\cA}} = 0 \ , \quad \cD_6 = \D
\ee
and
\begin{align} \tilde{\S} &= \frac{1}{5!} \rd z^{\hat{E}} \wedge \rd z^{\hat{D}} \wedge \rd z^{\hat{C}} \wedge \rd z^{\hat{B}} \wedge \rd z^{\hat{A}} \S_{\hat{A} \hat{B} \hat{C} \hat{D} \hat{E}} \ , \non\\
\Phi &= \frac{1}{4!} \rd z^{\hat{D}} \wedge \rd z^{\hat{C}} \wedge \rd z^{\hat{B}} \wedge \rd z^{\hat{A}} \S_{6 \hat{A} \hat{B} \hat{C} \hat{D}} \ .
\end{align}

We now impose simple constraints on the lowest mass dimension components
\begin{align}
\S_{\underline{\hat{\a}} \underline{\hat{\b}} \underline{\hat{\g}} \underline{\hat{\d}} \underline{\hat{\e}}} 
&= \S_{\hat{a} \underline{\hat{\a}} \underline{\hat{\b}} \underline{\hat{\g}} \underline{\hat{\d}}}
= \S_{\hat{a} \hat{b}  \underline{\hat{\a}} \underline{\hat{\b}} \underline{\hat{\g}}} 
= 
\S_{6  \underline{\hat{\a}} \underline{\hat{\b}} \underline{\hat{\g}} \underline{\hat{\d}}} 
= 
\S_{6 \hat{a} \underline{\hat{\b}} \underline{\hat{\g}} \underline{\hat{\d}}} = 0 \ , \non\\
\S_{6 \hat{a} \hat{b} \underline{\hat{\a}} \underline{\hat{\b}}} &= 4 \ri (\S_{\hat{a} \hat{b}})_{\hat{\a} \hat{\b}} L^{ij} \ ,
\end{align}
and analyze eq. \eqref{BISigma}. The remaining components are fixed as follows:
\begin{align}
\S_{\hat{a} \hat{b} \hat{c} \underline{\hat{\a}} \underline{\hat{\b}}} &=  2 \ri \eps_{\hat{a} \hat{b} \hat{c} \hat{d} \hat{e}} (\S^{\hat{d} \hat{e}})_{\hat{\a} \hat{\b}} \cW L^{ij} \ , \non\\
\S_{6 \hat{a} \hat{b} \hat{c} \underline{\hat{\a}}} &= - \frac{1}{3} \eps_{\hat{a} \hat{b} \hat{c} \hat{d} \hat{e}} (\S^{\hat{d} \hat{e}})_{\hat{\a}}{}^{\hat{\b}} \cD_{\hat{\b} j}L^{ji} \ , \non\\
\S_{\hat{a} \hat{b} \hat{c} \hat{d} \underline{\hat{\a}}} &= - \frac{1}{3} \eps_{\hat{a} \hat{b} \hat{c} \hat{d} \hat{e}} (\G^{\hat{e}})_{\hat{\a}}{}^{\hat{\b}} (\cW \cD_{\hat{\b} j}L^{ji} + 3 \cD_{\hat{\b} j} \cW L^{ji}) \ , \non\\
\S_{6 \hat{a} \hat{b} \hat{c} \hat{d}} &= \frac{\ri}{24} \eps_{\hat{a} \hat{b} \hat{c} \hat{d} \hat{e}} (\G^{\hat{e}})^{\hat{\a} \hat{\b}} \cD_{\hat{\a}}^i \cD_{\hat{\b}}^j L_{ij} \ , \non\\
\S_{\hat{a} \hat{b} \hat{c} \hat{d} \hat{e}} &= \frac{\ri}{24} \eps_{\hat{a} \hat{b} \hat{c} \hat{d} \hat{e}} (\cW \cD^{\hat{\g} i} \cD_{\hat{\g}}^j L_{ij} + 3 \cD^{\hat{\g} i} \cD_{\hat{\g}}^j \cW L_{ij} 
+ 8 \cD^{\hat{\g} i} \cW \cD_{\hat{\g}}^j L_{ij} ) \ ,
\end{align}
where $L^{ij}$ satisfies the constraint for the linear supermultiplet
\be \cD_{\hat{\a}}^{(i} L^{jk)} = 0 \ . \label{constLinear}
\ee

In the above we did not assume anywhere that $L^{ij}$ is annihilated by the central charge. However, if $L^{ij}$ is inert under the central charge, $\D L^{ij}=0$, we have
\be \rd \Phi = 0
\ee
and $L^{ij}$ becomes a gauge three-form supermultiplet, also known as the $\cO(2)$ supermultiplet.


\subsubsection{Action principle}


Making use of the components $\S_{\hat{\cA} \hat{\cB} \hat{\cC} \hat{\cD} \hat{\cE}}$ one can construct a closed five-form.
The appropriate closed form is simply given by
\be
J = \tilde{\S} - \cV \wedge \Phi \ .
\ee
All that one must check is closure,
\be \rd J = \rd \tilde{\S} - \cV \wedge \rd \Phi - \rd \cV \wedge \Phi = \cD \tilde{\S} - \cV \wedge \D \tilde{\S} - \cV \wedge \cD \Phi - \cF \wedge \Phi = 0 \ ,
\ee
and the transformation law under central charge transformations,
\begin{align}
\d_{\L} J &= \d_{\L} \S + \d_{\L} \cV \wedge \Phi + \cV \wedge \d_{\L} \Phi \non\\
&= \L \D \S - \rd \L \wedge \Phi + \cV \wedge (\L \D \Phi) = \rd (\L \D \Phi) \ .
\end{align}
In components we have
\bsubeq
\be J_{\hat{a}\hat{b}\hat{c}\hat{d}\hat{e}} = \tilde{\S}_{\hat{a}\hat{b}\hat{c}\hat{d}\hat{e}} - 5 \cV_{[\hat{a}} \Phi_{\hat{b}\hat{c}\hat{d}\hat{e}]} \ ,
\ee
which gives
\be
{}^*J 
= \frac{1}{5!} \eps^{\hat{a}\hat{b}\hat{c}\hat{d}\hat{e}} \tilde{\S}_{\hat{a}\hat{b}\hat{c}\hat{d}\hat{e}} - \frac{1}{4!} \eps^{\hat{a}\hat{b}\hat{c}\hat{d}\hat{e}} \cV_{\hat{a}} \Phi_{\hat{b}\hat{c}\hat{d}\hat{e}} \ .
\ee
\esubeq
The action is then
\begin{align} \label{LinearCompAction}
S &= - \frac{\ri}{24} \int \rd^5x \Big(\cW \cD^{\hat{\g} i} \cD_{\hat{\g}}^j L_{ij} 
+ 3 \cD^{\hat{\g} i} \cD_{\hat{\g}}^j \cW L_{ij} + 8 \cD^{\hat{\g} i} \cW \cD_{\hat{\g}}^j L_{ij} 
+ \cV_{\hat{a}} (\cD^i \G^{\hat{a}} \cD^j) L_{ij}\Big) \Big|  \non\\
&= - \frac{1}{2} \int \rd^5x \Big(\varphi G
+ X^{ij} \ell_{ij} 
+ 2 \Psi^{\hat{\g} k} \c_{\hat{\g} k}
- 2 \cV_{\hat{a}} \phi^{\hat{a}}\Big) \ ,
\end{align}
where the component fields of $\cW$ are defined as in eq. \eqref{VMcompDef} and we have defined the component fields 
of the linear supermultiplet as follows:
\bsubeq
\begin{align} \ell^{ij} &:= L^{ij}| \ , \quad \c_\hal^i := \frac{1}{3} \cD_{\hal j} L^{ij}| \ , \quad G := \frac{\ri}{12} \cD^{\hga i} \cD_\hga^j L_{ij}| \ , \\
\phi^\ha &:= \frac{\ri}{24} (\G^\ha)^{\hal\hbe} \cD_\hal^i \cD_\hbe^j L_{ij}| = \Phi^\ha| \ , \quad \Phi_{\ha\hb\hc\hd} = \eps_{\ha\hb\hc\hd\he} \Phi^\he \ .
\end{align}
\esubeq

The supersymmetry transformations for the linear supermultiplet 
 follow from the constraint \eqref{constLinear} and are found to be
\bsubeq \label{SUSYlinear}
\begin{align} \d_\xi \ell^{ij} &= - 2 \xi^{\hal (i} \c_\hal^{j)} \ , \\
\d_\xi \c_\hal^i &= - \frac{\ri}{2} \xi_\hal^i G + \ri \xi^{\hbe i} \phi_{\hal\hbe} + \ri \xi^\hbe_j \cD_{\hbe\hal} \ell^{ij} \ , \\
\d_\xi G &= - 2 \xi^\hal_i \cD_\hal{}^\hbe \c_\hbe^i - 2 \ri \xi^\hal_i \Psi_{\hal j} \D \ell^{ij} \ , \\
\d_\xi \phi_\ha &= 2 \xi^\hal_i (\S_{\ha\hb})_\hal{}^\hbe \cD^\hb \c_\hbe^i - \ri \xi^\hal_i (\G_\ha)_\hal{}^\hbe \Psi_{\hbe j} \D \ell^{ij} 
- \xi^\hal_i (\G_\ha)_\hal{}^\hbe \varphi \D \c_\hbe^i \ . \label{SUSYPhia}
\end{align}
\esubeq
The action \eqref{LinearCompAction} and the supersymmetry transformations \eqref{SUSYlinear} agree with those given in \cite{KO-I}. These 
results hold for the linear multiplet both with or without central charge.
 
It is worth noting that checking invariance of the component action \eqref{LinearCompAction} under the central charge is nontrivial and requires having to 
derive some nontrivial identities. However, within the superform approach invariance follows much more easily. Furthermore, the 
superform formulation for the linear supermultiplet tells us more than just the action. For instance, taking the 
component projection of the Bianchi identity
\be 5 \cD_{[\ha} \Phi_{\hb\hc\hd\he\}} = \D \tilde{\S}_{\ha\hb\hc\hd\he} \ ,
\ee
gives the differential constraint on the component field $\phi_\ha$
\be 2 \cD^\ha \phi_\ha = \D (\varphi G + X^{ij} \ell_{ij} + 2 \psi^{\hga k} \c_{\hga k}) \ .
\ee
The supersymmetry transformations are also encoded in the Bianchi identities \eqref{BISigma}. This provides an efficient means of 
computing some of the supersymmetry transformations. In particular, the supersymmetry transformation of $\phi_\ha$, eq. \eqref{SUSYPhia}, follows directly from 
the component projection of the Bianchi identity
\be \cD_\hal^i \Phi_{\ha\hb\hc\hd} = - 4 \cD_{[\ha} \Phi_{\hb\hc\hd ]}{}_\hal^i + \D \tilde{\S}_{\ha\hb\hc\hd}{}_\hal^i \ .
\ee

Using the action for the linear supermultiplet, one can derive the Abelian Chern-Simons action by taking \cite{KL}
\be L^{ij} = \ri \cD^{\hat{\g} (i} W \cD_{\hat{\g}}^{j)} W + \frac{\ri}{2} W \cD^{\hat{\g} (i} \cD_{\hat{\g}}^{j)} W \ .
\ee
Using the above choice of $L^{ij}$ and the action principle for the linear supermultiplet one derives (after removing a total derivative from the Lagrangian and dividing out 
an irrelevant factor of $6$) the Abelian action \eqref{abelianVMaction}.


\subsection{Gauge two-form supermultiplet}

We have seen how to derive the Abelian Chern-Simons action both by constructing a curvature induced form and by making use of 
the linear supermultiplet. The vector supermultiplet turns out to be dual to a gauge two-form supermultiplet, which possesses an intrinsic 
central charge and may be coupled to additional vector supermultiplets via Chern-Simons terms. The supermultiplet is also called the gauge tensor
multiplet or small tensor multiplet in \cite{KO-III}.\footnote{On-shell tensor 
multiplets in 5D gauged supergravity were introduced by G\"unaydin and Zagermann \cite{GZ} (see also \cite{Ceresole}) 
as a generalization of the earlier work by G\"unaydin, Sierra and Townsend
\cite{GST} on 5D supergravity-matter systems with vector supermultiplets.}
In superspace, it is described, 
similar to the 4D $\cN=2$ vector-tensor supermultiplet \cite{DIKST},   
by a constrained real  superfield $L$ coupled to the vector supermultiplet gauging the central charge
\cite{Kuzenko:5D}.  
 In this subsection, we will turn to deriving a 
superform formulation for this supermultiplet.

We start with the superspace setting of subsection \ref{gauging} in which the central charge is gauged by a vector supermultiplet $\cW$. However, 
we will also include coupling to an additional Yang-Mills supermultiplet $W$ (see subsection \ref{YM}). Therefore in this subsection we will make 
use of covariant derivatives which include both gauge connections\footnote{The central charge commutes with the Yang-Mills gauge group.}
\be \cD = \rd + \cV \D + \ri V \ , \quad \cD_{\hat{A}} = D_{\hat{A}} + \cV_{\hat{A}} \D + \ri V_{\hat{A}} \ .
\ee

We introduce 
a gauge two-form, $B = \hf E^B E^A B_{AB}$ and its three-form 
field strength $H$ defined by\footnote{Both $B$ and $H$ are Yang-Mills 
singlets.}
 \be H := \cD B - \tr \big( V \wedge F + \frac{\ri}{3} V \wedge V \wedge V \big) \ ,
\ee
where $V$ and $F$ is the Yang-Mills connection and field strength corresponding to the superfield $W$.\footnote{The special 
case of $n$ Abelian vector supermultiplets may be obtained by taking $\tr(V \wedge F) \rightarrow \eta_{IJ} V^I F^J$, where 
$\eta$ is a symmetric, $\eta_{IJ} = \eta_{JI}$, coupling constant and $V^I$ and $F^I$ are the gauge connections and field strengths of the Abelian vector supermultiplets.} 
Here we do not assume $B$ to be annihilated by the central charge. The (infinitesimal) transformation law for the system of superforms is
\begin{align}
\d \cV &= - \rd \L \ , \quad \D \L = 0 \ , \non\\
\d V &= - \rd \t \ , \quad \D \t = 0 \ , \non\\
\d B &= \L \D B - \tr(\t \wedge \rd V) + \rd \G \ , \quad \D \G = 0 \ ,
\end{align}
where $\L$, $\t$ and $\G$ generate the gauge transformations of $\cV$, $V$ and $B$ respectively. The field strength $H$ transforms covariantly
\be \d H = \L \D H
\ee
and satisfies the Bianchi identity
\be \cD H = \cF \wedge \D B - \tr (F \wedge F) \ . \label{VTBI1}
\ee

Using the notation that was introduced in subsection \ref{SFlinear}, it is possible to extend the Bianchi identity by introducing an additional 
bosonic index, $\hat{\cA} = (\hat{A} , 6)$. To do this we first note that we also have the additional superform equation
\be \D H = \cD (\D B) \ . \label{VTBI2}
\ee
We then extend the Bianchi identity \eqref{VTBI1} and the additional equation \eqref{VTBI2} to
\be \cD_{[\hat{\cA}} H_{\hat{\cB} \hat{\cC} \hat{\cD} \}} - \frac{3}{2} T_{[\hat{\cA} \hat{\cB}}{}^{\hat{\cE}} H_{|\hat{\cE}| \hat{\cC} \hat{\cD} \}} 
+ \frac{3}{2} \tr(F_{[\hat{\cA} \hat{\cB}} F_{\hat{\cC} \hat{\cD} \}}) = 0 \ , \label{GBI1F}
\ee
where we have defined
\bsubeq \label{VTID}
\begin{align} H_{6 \hat{A} \hat{B}} &:= \D B_{\hat{A} \hat{B}} \ , \quad F_{6 \hat{\cA}} = F_{\hat{\cA} 6} = 0 \ , \\
T_{\hat{A} \hat{B}}{}^6 &:= \cF_{\hat{A} \hat{B}} \ , \quad T_{\hat{A} 6}{}^{\hat{\cB}} = T_{6 \hat{A}}{}^{\hat{\cB}} = 0 \ , \quad \cD_6 := \D \ .
\end{align}
\esubeq

We now impose simple constraints on the lowest components of $H_{\hat{\cA} \hat{\cB} \hat{\cC}}$
\be H_{\underline{\hat{\a}}\underline{\hat{\b}}\underline{\hat{\g}}} = 0 \ , \quad H_{6 \underline{\hat{\a}} \underline{\hat{\b}}} = - 2 \ri \eps^{ij} \eps_{\hat{\a} \hat{\b}} L \ . \label{2FConst}
\ee
The remaining components of $H_{\hat{\cA} \hat{\cB} \hat{\cC}}$ can be found by analyzing eq. \eqref{GBI1F} subject to the constraints \eqref{2FConst} and the identifications \eqref{VTID}. 
They are found to be:
\bsubeq
\begin{align} H_{\hat{a} \underline{\hat{\b}} \underline{\hat{\g}}} &= - 2 \ri \eps^{jk} \big(\G_{\hat{a}})_{\hat{\b} \hat{\g}} (\cW L- \tr(W^2)\big) \ , \\
H_{6 \hat{a}}{}_{\hat{\b}}^j &= (\G_{\hat{a}})_{\hat{\b}}{}^{\hat{\g}} \cD_{\hat{\g}}^j L \ , \\
H_{\hat{a} \hat{b}}{}_{\hat{\g}}^k &=  2 (\S_{\hat{a} \hat{b}})_{\hat{\g}}{}^{\hat{\d}} \cD_{\hat{\d}}^k (\cW L - \tr (W^2 )) \ , \\
H_{6 \hat{a} \hat{b}} &= \frac{\ri}{4} (\S_{\hat{a}\hat{b}})^{\hat{\a} \hat{\b}} \cD^k_{\hat{\a}} \cD_{\hat{\b} k} L \ , \non\\
H_{\hat{a} \hat{b} \hat{c}} &= - \frac{\ri}{8} \eps_{\hat{a} \hat{b} \hat{c} \hat{d} \hat{e}} (\S^{\hat{d} \hat{e}})^{\hat{\a} \hat{\b}} \Big( \cD_{\hat{\a}}^k \cD_{\hat{\b} k} \big(\cW L - \tr(W^2) \big) \non\\
&+ 2 \cD_{\hat{\a}}^k \cW \cD_{\hat{\b} k} L - 2 \tr(\cD_{\hat{\a}}^k W \cD_{\hat{\b} k} W)\Big) \ ,
\end{align}
\esubeq
where $L$ satisfies the constraints
\bsubeq \label{2FORMCONST}
\begin{align} \cD_{\hat{\a}}^{(i} \cD_{\hat{\b}}^{j)} L &= \frac{1}{4} \eps_{\hat{\a} \hat{\b}} \cD^{\hat{\g} (i} \cD_{\hat{\g}}^{j)} L \ , \label{3.31a} \\
\cD^{\hat{\g} (i} \cD_{\hat{\g}}^{j)} \big(\cW L - \tr(W^2) \big) &= - 2 \cD^{\hat{\g} (i} \cW \cD_{\hat{\g}}^{j)} L + 2 \tr( \cD^{\hat{\g} (i} W \cD_{\hat{\g}}^{j)} W) \label{3.31b} \ .
\end{align}
\esubeq
The constraints derived from the geometry precisely agree with those in \cite{Kuzenko:5D}. The remarkable feature of this analysis is that it highlights 
how the  constraints \eqref{2FORMCONST} follow from requiring the presence of a two-form and simple constraints on its field-strength.

The corresponding superfield Lagrangian may be taken as \cite{Kuzenko:5D} (formally the same as that of a vector supermultiplet)
\be 
L^{ij} = \frac{\ri}{2} \big( 2 \cD^{\hat{\a} (i} L \cD_{\hat{\a}}^{j)} L + L \cD^{\hat{\a} (i} \cD_{\hat{\a}}^{j)} L \big) \ .
\label{3.32}
\ee
The equation of motion for this model proves to be  $\D L = 0$.

The off-shell component action for the gauge two-form supermultiplet (in supergravity) together with its Chern-Simons couplings was constructed in \cite{KO-III}. The 
formulation of the Chern-Simons couplings was inspired by the general form of vector-tensor supermultiplet couplings in the superconformal framework 
\cite{BdeWGHVvanP}.

\subsection{Large tensor supermultiplet}\label{largeTensorMultiplet}

In \cite{KO-III} it was discovered that there also exists the large tensor supermultiplet, which consists of $16 \ ({\rm boson}) + 16 \ ({\rm fermion})$ component fields. The large 
tensor supermultiplet can also be seen to naturally originate in superspace. It may be viewed as a generalization of the gauge two-form supermultiplet in which the constraints 
\eqref{2FORMCONST} are weakened. To show this let $\cL$ be a superfield constrained in the same way as eq. \eqref{3.31a},
\be \cD_{\hat{\a}}^{(i} \cD_{\hat{\b}}^{j)} \cL 
= \frac{1}{4} \eps_{\hat{\a} \hat{\b}} \cD^{\hat{\g} (i} \cD_{\hat{\g}}^{j)} \cL \label{3.33} \ .
\ee
Requiring only the above constraint, it is possible to show that consistency requires us to have
\cite{Kuzenko:5D} 
\bea 
0&=& \D \Big\{ 
\cD^{\hat{\g} (i} \cD_{\hat{\g}}^{j)} (\cW  \cL )  + 2 \cD^{\hat{\g} (i} \cW \cD_{\hat{\g}}^{j)}  
\cL \Big\}     
 \non \\
&=&\cD^{\hat{\g} (i} \cD_{\hat{\g}}^{j)} (\cW \D \cL )  
+ 2 \cD^{\hat{\g} (i} \cW \cD_{\hat{\g}}^{j)} \D \cL  \ , ~~~~
\label{3.34}
\eea
which is automatically satisfied for the gauge two-form supermultiplet. 
Here we will take eq. \eqref{3.34} as a second constraint on $\cL$. 
The constraints \eqref{3.33} and \eqref{3.34} allow us to construct a superform 
framework describing the large tensor supermultiplet.

We begin by introducing a two-form $\cB$, transforming homogeneously under the local central charge transformations
\be \d \cB = \L \D \cB\ ,
\ee
and an associated three form $\cH$
\be \cH = \cD \cB \ . \label{hatHSE}
\ee
Imposing the constraints
\be \cH_{\underline{\hat{\a}} \underline{\hat{\b}} \underline{\hat{\g}}} = 0 \ , \quad \cH_{6 \underline{\hat{\a}} \underline{\hat{\b}}} = - 2 \ri \eps^{ij} \eps_{\hat{\a}\hat{\b}} \D \cL
\ee
and solving the Bianchi identities yields the components of $\cH$:
\bsubeq
\begin{align} \cH_{\hat{a} \underline{\hat{\b}} \underline{\hat{\g}}} &= - 2 \ri \eps^{jk} \big(\G_{\hat{a}})_{\hat{\b} \hat{\g}} \cW \D \cL \ , \\
\cH_{6 \hat{a}}{}_{\hat{\b}}^j &= (\G_{\hat{a}})_{\hat{\b}}{}^{\hat{\g}} \cD_{\hat{\g}}^j \D \cL \ , \\
\cH_{\hat{a} \hat{b}}{}_{\hat{\g}}^k &
=  2 (\S_{\hat{a} \hat{b}})_{\hat{\g}}{}^{\hat{\d}} \cD_{\hat{\d}}^k (\cW \D \cL) \ , \\
\cH_{6 \hat{a} \hat{b}} &= \frac{\ri}{4} (\S_{\hat{a}\hat{b}})^{\hat{\a} \hat{\b}} \cD^k_{\hat{\a}} \cD_{\hat{\b} k} \D \cL \ , \non\\
\cH_{\hat{a} \hat{b} \hat{c}} &= - \frac{\ri}{8} \eps_{\hat{a} \hat{b} \hat{c} \hat{d} \hat{e}} (\S^{\hat{d} \hat{e}})^{\hat{\a} \hat{\b}} \Big( \cD_{\hat{\a}}^k \cD_{\hat{\b} k} \big(\cW \D \cL\big) 
+ 2 \cD_{\hat{\a}}^k \cW \cD_{\hat{\b} k}  \D \cL \Big) \ ,
\end{align}
\esubeq
where $\cL$ is constrained by eqs. \eqref{3.33} and \eqref{3.34} and $\cH_{6 \hat{A} \hat{B}} = \D \cB_{\hat{A} \hat{B}}$. There are still too many component fields and to eliminate them we 
impose the constraint on $\cB$
\be \cB_{\hat{\a}}^i{}_{\hat{\b}}^j = - 2 \ri \eps^{ij} \eps_{\hat{\a} \hat{\b}} \cL \ ,
\ee
which fixes the remaining components via eq. \eqref{hatHSE} as
\be \cB_{\hat{a}}{}_{\hat{\b}}^j = (\G_{\hat{a}})_{\hat{\b}}{}^{\hat{\g}} \cD_{\hat{\g}}^j \cL \ , 
\quad \cB_{\hat{a} \hat{b}} 
= \frac{\ri}{4} (\S_{\hat{a}\hat{b}})^{\hat{\a} \hat{\b}} \cD^k_{\hat{\a}} \cD_{\hat{\b} k} \cL \ .
\ee
At the highest dimension eq. \eqref{hatHSE} gives
\be
3 \cD_{[\hat{a}} \cB_{\hat{b} \hat{c}]} = - \frac{\ri}{8} \eps_{\hat{a} \hat{b} \hat{c} \hat{d} \hat{e}} (\S^{\hat{d} \hat{e}})^{\hat{\a} \hat{\b}} \D \Big( \cD_{\hat{\a}}^k \cD_{\hat{\b} k} \big(\cW \cL\big) 
+ 2 \cD_{\hat{\a}}^k \cW \cD_{\hat{\b} k} \cL \Big)  \ . \label{firstKO}
\ee
The conditions  \eqref{3.34} and \eqref{firstKO}  correspond to the ones imposed 
in \cite{KO-III} from requiring 
closure of the supersymmetry transformations. In contrast with the 
gauge two-form supermultiplet, which was based on the stronger constraints \eqref{2FORMCONST}, the component fields of the 
large tensor supermultiplet
\be \D \cD_\a^i \cL | \ , \quad \D^2 \cL |
\ee
are no longer composite. We should remark that the above constraints can naturally be generalized to include couplings to the 
Yang-Mills supermultiplet.

We can construct an action for an even number of large tensor supermultiplets $\cL^I$. 
To do so we make use of the superfield Lagrangian
\be \cL^{ij} = \cL_{\rm kin}^{ij} + \cL_{\rm mass}^{ij} \ ,
\ee
where
\bsubeq
\begin{align} \cL_{\rm mass}^{ij} &= \frac{\ri}{2} m_{IJ} \big( 2 \cD^{\hat{\a} (i} \cL^I \cD_{\hat{\a}}^{j)} \cL^J + \cL^I \cD^{\hat{\a} (i} \cD_{\hat{\a}}^{j)} \cL^J \big) \ , \quad m_{IJ} = m_{JI} \ , \\
\cL_{\rm kin}^{ij} &= \frac{\ri}{4} k_{IJ} \big( 2 \cD^{\hat{\a} (i} \cL^I \overleftrightarrow{\D} \cD_{\hat{\a}}^{j)} \cL^J 
+ \cL^I \overleftrightarrow{\D} \cD^{\hat{\a} (i} \cD_{\hat{\a}}^{j)} \cL^J \big) \ , \quad k_{IJ} = - k_{JI} \ .
\end{align}
\esubeq
The constant matrices $m_{IJ}$ and $k_{IJ}$ are assumed to be nonsingular. 
The Lagrangian $\cL^{ij}$ may be seen to be a linear supermultiplet.
The component action in supergravity is given in \cite{KO-III}.

On-shell each large tensor supermultiplet describes 4 + 4 degrees of freedom \cite{KO}. The equations of motion for the large-tensor supermultiplets are given by the 
superfield constraint
\be k_{IJ} \D \cL^J + m_{IJ} \cL^J = 0 \ . \label{EOM}
\ee
Under the above constraint \eqref{firstKO} becomes a duality condition on $\cB$, 
\be
\hf k_{IJ} \eps^{\hat{a} \hat{b} \hat{c} \hat{d} \hat{e}}  \cD_{\hat{c}} \cB_{\hat{d} \hat{e}}^J = - m_{IJ} \Big( \cW \cB^{J \hat{a}\hat{b}} + F^{\hat{a}\hat{b}} \cL^J
+ \ri (\S^{\hat{a} \hat{b}})^{\hat{\a} \hat{\b}} \cD_{\hat{\a}}^k \cW  \cD_{\hat{\b} k} \cL^J \Big) \ . \label{firstKO2}
\ee
Furthermore, the 16 + 16 independent component fields
\be \cL| \ , \quad \cD_{\hat{\a}}^i \cL| \ , \quad \D \cL| \ , \quad \cB_{\hat{a}\hat{b}}| \ , \quad \cD^{\hat{\g} (i} \cD_{\hat{\g}}^{j)} \cL| \ , \quad \D \cD_{\hat{\a}}^i \cL| \ , \quad \D^2 \cL| \ ,
\ee
reduce to\footnote{The component field $\cD^{\hat{\g} (i} \cD_{\hat{\g}}^{j)} \cL|$ is 
composite as a result of eqs. \eqref{3.34} and \eqref{EOM}.}
\be \cL| \ , \quad \cD_{\hat{\a}}^i \cL| \ , \quad \cB_{\hat{a}\hat{b}}| \ .
\ee
These components correspond to only $4+4$ degrees of freedom. To see this, we note that the self-duality condition \eqref{firstKO2} implies 
that $\cB_{\hat{a}\hat{b}}|$ now possesses only 3 
degrees of freedom. Therefore we have $3+1 = 4$ bosonic degrees of freedom. The remaining 
component field $\cD_{\hat{\a}}^i \cL|$ contributes to the remaining 4 fermionic degrees of freedom.


\section{Discussion} \label{conclusion}

The closed-form expression for the non-Abelian SCS action in 5D $\cN=1$ superspace
is one of the main results of this paper.  The component action was constructed
by Kugo and Ohashi more than ten years ago \cite{KO}. However, our work has provided
the first systematic, unambiguous and purely geometric derivation of this action.  
Our construction can readily be generalized to the locally supersymmetric case 
by making use of the superspace formulation for 
5D $\cN=1$ conformal supergravity \cite{KT-M08}. 
Moreover, we believe our construction makes it it possible to address another 
long-standing problem -- to formulate the 5D $\cN=1$ non-Abelian SCS action 
in terms of 4D $\cN=1$ superfields. For this one has to use the relations 
\eqref{md5Comp} and \eqref{2.28a}
in conjunction with the formalism of reduced superspace introduced in \cite{KL}. 
We hope to elaborate on this issue elsewhere. 

The idea of generalizing the gauge two-form supermultiplet in the way described in subsection \eqref{largeTensorMultiplet} may have an immediate application 
for the vector-tensor supermultiplet in four-dimensions. 
To see this, we first recall that in superspace 
 the vector-tensor supermultiplet with  gauged central charge
$\mathbb{L}$ satisfies the constraint\footnote{There are also 
additional constraints which are not important here.}
\be \cD_\a^{(i} \bar{\cD}_\ad^{j)} \mathbb{L} = 0 \ . \label{4.1} 
\ee
The above constraint can only be consistent if the following additional constraint is imposed \cite{DIKST}
\bea 
0&=&\D \Big(\cD^{\a (i} \cD_\a^{j)} (\cW \mathbb{L}) + \bar{\cD}_\ad^{(i} \bar{\cD}^{\ad j)} (\bar{\cW} \mathbb{L}) - \mathbb{L} \cD^{\a (i} \cD_\a^{j)} \cW \Big) \non \\
&=& \cD^{\a (i} \cD_\a^{j)} (\cW \D \mathbb{L}) + \bar{\cD}_\ad^{(i} \bar{\cD}^{\ad j)} (\bar{\cW} \D \mathbb{L}) -\cD^{\a (i} \cD_\a^{j)} \cW  \D \mathbb{L} 
\label{4.2} \ ,
\eea
where $\cW$ is the chiral field strength of the 4D $\cN=2$ central charge vector supermultiplet, 
\be \cD^{\a (i} \cD_\a^{j)} \cW = \bar{\cD}_\ad^{(i} \bar{\cD}^{\ad j)} \bar{\cW} \ .
\ee
Although stronger constraints are usually chosen for $\mathbb{L}$, our analysis of the large tensor supermultiplet suggests that we could instead choose eq. \eqref{4.2} as a 
second constraint and look for a consistent superform formulation. Furthermore, a similar possibility exists for the variant vector-tensor supermultiplet 
\cite{Theis1, Theis2, Novak1}.\footnote{The analogue of \eqref{4.2} for the variant vector-tensor supermultiplet may be found in \cite{Novak1}.} Whether the more 
general constraints will lead to consistent supermultiplets is still an open problem.
\\


\noindent
{\bf Acknowledgements:}\\
We are grateful to Daniel Butter for useful discussions. 
SMK  is grateful to INFN, Padova Section and 
the Department of Physics and Astronomy ``Galileo Galilei'' at the 
University of Padova for kind hospitality at the final stage of this project. 
This work was supported in part by the Australian Research Council,
project No. DP1096372.  
The work of JN was also supported in part by the Australian Research Council's Discovery Early Career 
Award (DECRA), project No. DE120101498.


\appendix

\section{Alternative gauging of the central charge} \label{appendixB}

In subsection \eqref{gauging} we made use of a vector supermultiplet to gauge the central charge. This requires the gauge potential $\cV$ to be 
inert under the action of the central charge, $\D \cV = 0$. However, it was shown in \cite{Theis1, Theis2} that in 4D it is possible to gauge the central 
charge with a gauge connection that is not annihilated by the central charge. To the best of our knowledge, the possibility of gauging the central charge 
with a gauge connection that is no longer inert under the central charge has never been properly analyzed in 5D. In this appendix, we follow an approach 
similar to that given in \cite{Novak1, Novak2}. We do not assume that the gauge one-form is annihilated by the central charge and analyze the possibilities 
under reasonable constraints.

We begin as in subsection \eqref{gauging} by introducing gauge covariant derivatives
\be \cD_{\hat{A}} = (\cD_{\hat{a}}, \cD_{\hat{\a}}^i) = D_{\hat{A}} + \cV_{\hat{A}} \D \ , \quad [\D, D_{\hat{A}}] = 0 \ ,
\ee
where $\cV_{\hat{A}}$ is a one-form gauge connection associated with the central charge $\D$ and $\D \cV_{\hat{A}} \neq 0$. Here the 
gauge transformation of the gauge connection $\cV_{\hat{A}}$ becomes
$\cV_{\hat{A}}$ to be
\be \d \cV_{\hat{A}} = - D_{\hat{A}} \L + \L \D \cV_{\hat{A}} \implies \d \cD_{\hat{A}} = [\L \D , \cD_{\hat{A}}] \ , 
\ee
where the gauge parameter is annihilated by the central charge, $\D \L = 0$.

The commutation relations for the gauged covariant derivatives are
\bsubeq
\begin{align}
[ \cD_{\hat{A}}, \cD_{\hat{B}} \} &= T_{\hat{A} \hat{B}}{}^{\hat{C}} \cD_{\hat{C}} + \cF_{\hat{A} \hat{B}} \D \ , \\
[\D , \cD_{\hat{A}}] &= \cF_{6 \hat{A}} \D \ ,
\end{align}
\esubeq
where we define the field strengths
\bsubeq
\begin{align}
\cF_{\hat{A} \hat{B}} &:= 2 \cD_{[\hat{A}} \cV_{\hat{B} \}} - T_{\hat{A} \hat{B}}{}^{\hat{C}} \cV_{\hat{C}} \ , \\
\cF_{6 \hat{A}} &:= \D \cV_{\hat{A}} \ .
\end{align}
\esubeq
Here the field strengths $\cF_{\hat{A} \hat{B}}$ and $\cF_{6 \hat{A}}$ are covariant with respect to gauge transformations of $\cV_{\hat{A}}$
\be \d \cF_{\hat{A} \hat{B}} = \L \D \cF_{\hat{A} \hat{B}} \ , \quad \d \cF_{6 \hat{A}} = \L \D \cF_{6 \hat{A}} \ .
\ee
The Bianchi identities satisfied by $\cF_{\hat{A} \hat{B}}$ and $\cF_{6 \hat{A}}$ can be combined into one equation by extending the indices to 
include an additional bosonic coordinate, $\hat{\cA} = (\hat{A}, 6)$. The extended object $\cF_{\hat{\cA}\hat{\cB}} = (\cF_{\hat{A}\hat{B}} , \cF_{6\hat{A}})$
satisfies the Bianchi identity
\be \cD_{[\hat{\cA}}\cF_{\hat{\cB} \hat{\cC} \} } - T_{[\hat{\cA} \hat{\cB}}{}^{\hat{\cD}} \cF_{\hat{\cD} \hat{\cC} \} } = 0 \ , \label{EFSBI}
\ee
where we have made the identifications
\be T_{\hat{A} \hat{B}}{}^6 = \cF_{\hat{A} \hat{B}} \ , \quad T_{6 \hat{\cA}}{}^6 = - T_{\hat{\cA} 6}{}^6 = \cF_{6 \hat{\cA}} \ , \quad T_{6 \hat{\cA}}{}^{\hat{B}} = - T_{\hat{\cA} 6}{}^{\hat{B}} = 0 \ .
\ee

We may now impose constraints on the field strength and analyze the consequences of the Bianchi identities \eqref{EFSBI}. We choose the 
simple constraint
\be \cF_{\underline{\hat{\a}} \underline{\hat{\b}}} = - 2 \ri \eps^{ij} \eps_{\hat{\a} \hat{\b}} M \ ,
\ee
where $M$ is initially assumed to be an unconstrained superfield. Analyzing the Bianchi identities yields the components
\bsubeq
\begin{align} \cF_{\hat{a}}{}_{\hat{\b}}^j &= (\G_{\hat{a}})_{\hat{\b}}{}^{\hat{\g}} (\cD_{\hat{\g}}^j M - M \cF_{6}{}_{\hat{\g}}^j) \ , \\
 \cF_{6 \hat{a}} &= \frac{\ri}{8} (\G_{\hat{a}})_{\hat{\a} \hat{\b}} \cD^{\hat{\a} k} \cF_{6}{}^{\hat{\b}}_k \ , \\
\cF_{\hat{a} \hat{b}} &= \frac{\ri}{4} (\S_{\hat{a} \hat{b}})^{\hat{\a} \hat{\b}} (\cD_{\hat{\a}}^k \cD_{\hat{\b} k} M + M \cD_{\hat{\a} k} \cF_{6}{}_{\hat{\b}}^k + 2 \cD_{\hat{\a} k} M \cF_{6}{}_{\hat{\a}}^k)
\end{align}
\esubeq
and the constraints
\bsubeq
\begin{align} 
\cD_{\hat{\g}}^k \cF_6{}_k^{\hat{\g}} &= - 8 \ri \D M \ , \label{A.10a} \\
\cD_{(\hat{\a}}^{(i} \cF_6{}_{\hat{\b})}^{j)} &= 0 \ , \label{A.10b} \\
\cD_{\hat{\a}}^{(i} \cD_{\hat{\b}}^{j)} M &= \frac{1}{4} \eps_{\hat{\a} \hat{\b}} \cD^{\hat{\g} (i} \cD_{\hat{\g}}^{j)} M 
- \hf \eps_{\hat{\a} \hat{\b}} \cD^{\hat{\g} (i} M \cF_{6}{}_{\hat{\g}}^{j)} - \frac{1}{4} \eps_{\hat{\a} \hat{\b}} M \cD^{\hat{\g} (i} \cF_6{}_{\hat{\g}}^{j)} \non\\
&\quad+ 2 \cD^{(i}_{[\hat{\a}} M \cF_{6}{}_{\hat{\b}]}^{j)} + M \cD^{(i}_{[\hat{\a}} \cF_{6}{}_{\hat{\b}]}^{j)} \ .
\end{align}
\esubeq

If we first assume that $\D M \neq 0$ and all components of $\cF_{\hat{A} \hat{B}}$ are expressible in terms of $M$ and its covariant derivatives 
then the constraints \eqref{A.10a} and \eqref{A.10b} are solved by
\be \cF_{6}{}_{\hat{\b}}^j = \cD_{\hat{\b}}^j \ln M \ .
\ee
Putting this expression into the last constraint gives the condition
\be \cD_{\hat{\a}}^{(i} M \cD_{\hat{\b}}^{j)} M = \frac{1}{4} \eps_{\hat{\a} \hat{\b}} \cD^{\hat{\g} (i} M \cD_{\hat{\g}}^{j)} M \ ,
\ee
which implies
\be \cD_{\hat{\a}}^{(i} M \cD_{\hat{\b}}^j M \cD_{\hat{\g}}^{k)} M = 0 \ .
\ee
The only sensible solution to the above constraint is
\be \cD_{\hat{\a}}^i M = 0 \ .
\ee
However, this constraint implies that $\D M = 0$, which is a contradiction.

Choosing $\D M = 0$ reduces $M$ to that of a vector supermultiplet
\be M = W \ , \quad \cF_{6}{}_{\hat{\b}}^j = 0
\ee
with components given by eqns. \eqref{VGCC} and \eqref{VGCCBI}.

The result of our analysis is in stark contrast to the situation in 4D. In 4D it was pointed out by Theis \cite{Theis1, Theis2} that it is possible to gauge the central charge with the 
use of a different supermultiplet whose novel feature is that its gauge one-form is not annihilated by the central charge. The supermultiplet was later generalized to supergravity 
in \cite{Novak1, Novak2} 
and called the variant vector-tensor supermultiplet. The component structure of the supermultiplet is similar to that of the vector supermultiplet, possessing both a one-form and 
a two-form gauge field.\footnote{In \cite{Theis1, Theis2} the variant vector-tensor supermultiplet was called the new non-linear vector-tensor supermultiplet.} However, our 
analogous analysis in 5D shows that (under the reasonable assumptions made) the only supermultiplet suitable to gauge the central charge is the vector supermultiplet.

\begin{footnotesize}

\end{footnotesize}

\end{document}